\documentclass[a4paper]{ESASPCS13Style} \usepackage{epsfig}

\begin{document}

\title{X-rays, pre-main sequence stars and planet formation}

\author{E.\,D.  Feigelson}

\institute{Department of Astronomy \& Astrophysics, Pennsylvania State University,
University Park PA 16802 USA}

\maketitle

\begin{abstract}

The study of magnetic activity of pre-main sequence (PMS) stars through their X-ray
emission is entering a mature phase.  We review here two recent developments.
First, we present some early findings from the Chandra Orion Ultradeep Project
(COUP) relating to the age-acti\-vity-rotation relations of PMS stars.  COUP is a
nearly continuous exposure of the Orion Nebula Cluster over 13 days and gives the
most comprehensive view to date of X-ray emission from the PMS population.  We find
that the activity-rotation connection that dominates main sequence activity behavior is
absent in PMS stars, although the activity-age relation is present as well as
links to stellar mass and volume.  

Second, we discuss investigations of effects of PMS X-rays on their cold
protoplanetary disks.  It is possible that the the ionization, heating and
spallogenic effects of high energy photons and particles produced in PMS stellar
flares have significant effects on disk dynamics, heating and chemistry.  X-ray
ionization may induce MHD turbulence in the disk which, in turn, may regulate
disk evolution and planet formation processes such as Jovian planet migration.  

\end{abstract}

\section{Introduction}

The surface activity of stars during their PMS phases are not easily studied.  The
nearest active regions of star formation are $\simeq 150$ pc distant and the stars
often suffer interstellar obscuration.  Many techniques useful for studying the
stellar surfaces of closer and unobscured stars -- such as Zeeman spectroscopic
measurements, Doppler imaging, ultraviolet chromospheric indicators, radio
continuum flaring -- are restricted to the closest and brightest PMS stars.
However, elevated X-ray emission with strong flaring is ubiquitous from the early
protostellar stages through the T Tauri phases across the Initial Mass Function,
and can penetrate through hundreds of magnitudes of visual absorption.

X-ray have thus become an important indicator for PMS surface activity, and studies
during the 1980s and 1990s established that late-type stars exhibit their highest
levels of magnetic activity during their PMS phases (Feigelson \& Montmerle 1999).
But important puzzles remained unanswered.  Do magnetic fields in fully convective
stars descending the Hayashi tracks arise from the same dynamo mechanisms as in
solar-type stars, where the fields are generated in the shear layer between
radiative and convective zones?  Are the X-ray flares commonly seen in PMS stars
generated in magnetic loops with both footprints in the stellar surface, as in the
Sun, or might they arise from reconnection of field lines extending out to the
protoplanetary disk?  How does magnetic activity evolve with stellar age?  Do the
high energy processes involved in flaring have a significant impact on the complex
physics of the disk including accretion, outflows, and condensation into solid
bodies such as planets?

The Chandra X-ray Observatory with its ACIS CCD imaging array, a NASA Great
Observatory launched in 1999, is particularly well-suited to PMS stellar studies.
Its arcsecond imaging and subarcsecond astrometry permits unambiguous associations
between X-ray sources and stellar counterparts, even in rich and crowded young
stellar clusters.  Wide binaries and multiple systems are often resolved on scales
down to tens of AU.  The low noise of its CCD detectors and excellent rejection of
cosmic ray events permits detection of sources with fluxes as low as 1 photon per
day.  The mirror surface, polished to nearly single-atom precision, efficiently
reflects photons with energies up to 10 keV (wavelength 1 \AA); photons above
$\simeq 3$ keV can penetrate even the densest molecular cloud.  Chandra's
sensitivity to magnetic activity rivals the sensitivity of 8-meter-class telescopes
to photospheric emission in the optical and near-infrared bands.  The ACIS detector
produces a photon list in four-dimensions:  right ascen sion, declination, photon
energy, and photon arrival time.  Spectral and temporal analysis provides a
considerable body of information about the magnetic activity, the line-of-sight
obscuration, and occasionally about the circumstellar environment.

\section{The Chandra Orion Ultradeep Project}

The Orion Nebula has historically been an extraordinarily fruitful site for the
study of star formation and early stellar evolution.  The \object{Orion Nebula
Cluster} (ONC) has $\simeq 2000$ members concentrated in $\simeq 10$\arcmin\/
(1.3 pc) region with the best characterized stellar Initial Mass Function
ranging from a $\simeq 38$ M$_\odot$ O6-7 star to $\sim 3$ M$_J$ brown dwarfs.  The OB
members, known as the Trapezium, illuminate Messier 42, the nearest and
brightest HII region in the sky.  It is a blister HII region on the near side of
Orion A, the nearest giant molecular cloud.  Protoplanetary disks silhouetted
against the bright nebula have been directly imaged by the Hubble Space
Telescope around several dozen ONC members.  An arcminute northwest of the
Trapezium and behind it in the molecular cloud lies OMC-1, the nearest massive
molecular cloud core where star formation is now taking place.  These include a
dense cluster of intermediate-mass stars associated with the Orion Hot Core and
the Becklin-Neugebauer Object producing a powerful uncollimated molecular
outflow, and protostars producing collimated Herbig-Haro outflows in the Orion
1-South core.

The rich, crowded and obscured stellar field of the Orion Nebula provides a perfect
complement to Chandra's capabilities.  Several instrument teams studied the region
in the beginning of the mission producing a series of valuable studies (e.g.
Schulz et al.\ 2001; Feigelson et al.\ 2002; Flaccomio et al.\ 2003a).  These
groups then joined forces and, together with Orion experts at other wavebands and
theoretical astrophysicists, proposed an extended observation known as the Chandra
Orion Ultradeep Project (COUP).  It consists of a 9.7-day ACIS exposure taken over
13.2 days, a nearly-continuous observation with gaps only during perigee of the
satellite's orbit around Earth.  This gives superb sensitivity of $\log L_x = 27.0$
erg/s ($0.5-8$ keV band), sufficient to detect the contemporary active Sun, and
uniquely long temporal coverage to study flaring for 1616 sources.  Nearly all of
these are PMS stars in the ONC or OMC-1 regions.

COUP has many scientific objectives including improving the census of ONC
members, studying plasma properties in powerful stellar flares, elucidating the
origins and evolution of PMS X-ray emission, finding new embedded stars,
studying the magnetic activity of subsamples such as OB stars, brown dwarfs, and
stars with proplyds (i.e.  ("PROtoPLanetarY Disks") .  We will also use the
findings to improve theoretical studies of the effects of high energy photons
and particles on disk processes and evolution.  The reader is encouraged to
examine the original papers, some appearing in a 2005 dedicated issue of the
{\it Astrophysical Journal Supplements}, for detailed and quantitative findings.

\section{Age-activity-rotation relationships}

Here we restrict our discussion of COUP results on one topic:  extending the
relationships between stellar activity, age and rotation into the PMS evolutionary
phases.  These relationships have been a foundation of our understanding of stellar
magnetic activity, and have long been discussed at Cool Star Workshops.  Schatzman
(1962) developed a model of rotational deceleration in main sequence stars based on
magnetized stellar winds and Wilson (1963) showed that stellar chromospheric
activity was stronger in ZAMS stars like Pleiads than in older disk field stars.
Together, these studies led to an elaboration of empirical links between age,
rotation and activity such as Skumanich's (1972) $R_{HK} \propto v_{rot} \propto
t^{-1/2}$ which provided a basis for extending solar-type dynamo models to
late-type stars.

In the modern formulation of this $\alpha-\Omega$ dynamo (Schrijver \& Zwaan 2000),
the magnetic field generation scales with the rotational shear between the
radiative and convective zones at the tachocline, which presumable scales with the
measurable surface rotational velocity.  Talks at later CSSS workshops extended the
activity-rotation relationship to post-MS stars in tidally-locked binaries (Gunn et
al.\ 1998) but found it has difficulty when applied to fully convective M dwarfs
(Valenti et al.\ 1998) and completely breaks down for L dwarfs (Mohanty \& Basri
2003).  It is not clear that either very cool dwarfs or fully convective PMS stars
can exhibit the $\alpha-\Omega$ dynamo which requires magnetic field generation at
the tachoclinal interface between the radiation and convective zones.  The
possibility of convective zone dynamos was discussed as early as the 2nd CSSS
workshop by Gilman (1981) and are being seriously considered today for these stars.

While the temporal decay of magnetic activity in main sequence late-type stars
between ages of $\sim 0.1$ to $\sim 1$ Gyr has long been well-established from
chromospheric and coronal studies of open clusters, and has recently been extended
to the $1-10$ Gyr epoch (G\"udel et al.\ 1997; Feigelson et al.\ 2004).  At the 7th
CSSS workshop, Simon (1992) argued that stellar activity was physically dependent
entirely on rotation with only an indirect relationship with age.  The situation
for PMS stars has again been unclear whether this argument extends into the PMS
regime where a monotonic rotational deceleration is replaced by a complicated
competition between gravitational contraction, internal viscous coupling, and
external magnetic coupling between the stellar surface and the protoplanetary disk
(Barnes 2003).

\begin{figure}[ht] \begin{center} \epsfig{file=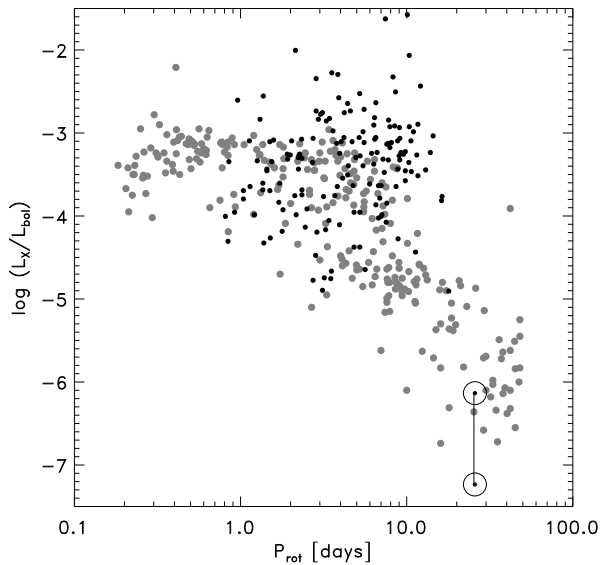, width=8cm}
  \end{center} \caption{X-ray emissivity in units of $\log L_x/L_{bol}$ plotted
  against photometric rotational periods for Orion Nebula stars.  The Orion stars
  are shown as black dots (time-averaged levels including flares),
  main sequence dwarfs are shown as grey dots (Pizzolato et al.\ 2003, Messina et 
  al.\ 2003), and the quiet-active Sun is shown as large circles.  Orion X-ray 
  luminosities are from the COUP study in the $0.5-8$ keV band corrected for 
  absorption (Getman et al.\ 2005), bolometric luminosities are from Hillenbrand 
  (1997) and rotation periods are from Herbst et al.\ (2002).  Preibisch et al., 
  in preparation.  } \end{figure}

Analysis of ROSAT observations of young stellar clusters suggested that magnetic
activity measured in the units $log L_x/L_{bol}$, essentially the magnetic
dissipation per unit area on the stellar surface, increases during the first few
Myr to the saturation level $\log L_x/L_{bol} \simeq -3$ as the protoplanetary disk
disappears, and declines on timescales of $10^7-10^8$ yr in a mass-dependent
fashion (Flaccomio et al.\ 2003b).  However, uncertainties arise from possible
incompleteness in the PMS samples, particularly for fainter weak-lined T Tauri
stars and from unresolved multiple systems.

\section{COUP results on age, activity and rotation}

The ONC probably provides the best possible sample for study of these questions.
Cluster membership has been carefully determined into the brown dwarf regime,
nearly a thousand stars have masses and ages estimated from placement on the
Hayashi tracks in the HR diagram, and several hundred members have surface
rotational periods measured from photometric modulation of cool star spots.  The
COUP findings on these issues are thus very promising avenues for understanding the
origins of magnetic activity in fully convective stars.  The work presented here in
preliminary form will appear in COUP studies led by Thomas Preibisch based on the
COUP source properties tabulated by Getman et al.\ (2005).

Figure 1 shows the X-ray emissivity $\log L_x/L_{bol}$ as a function of
rotational period for COUP Orion stars compared to main sequence stars.  Note that
the COUP observation is so sensitive that there are virtually no X-ray
nondetections in the sample.  No obvious association between X-ray emissivity and
rotation is seen.  Averaged over the Initial Mass Function, T Tauri X-ray levels
are typically slightly below the main sequence saturation level.  But other plots
of X-ray emission against other variables do show strong associations:  it scales
with stellar mass and functions of stellar radius including surface area and
internal volume.  One of these relations, plotting X-ray luminosity against stellar
volume, is shown in Figure 2.

Similar results were obtained from earlier ~ Chandra studies of the ONC but no
consensus on interpretation has emerged (Feigelson et al.\ 2003, Flaccomio et al.\
2003a, Stassun et al.\ 2004).  One one hand, a standard $\alpha-\Omega$ dynamo
scenario can be applied.  The displacement of the Orion stars below the saturated
level seen in MS stars in Figure 1 can be explained by a difference in the
convective zone depth and turnover time.  When the Rossby number rather than
rotational period is plotted, the Orion stars fall into the supersaturated regime
where $\log L_x/L_{bol} \simeq -3.5$ to $-4.0$ is typical.  The dependence on mass
may be unexplained but is also seen in a mass-dependence of the saturation level in
main sequence stars (Pizzolato et al.\ 2003).

\begin{figure}[ht] \begin{center} \epsfig{file=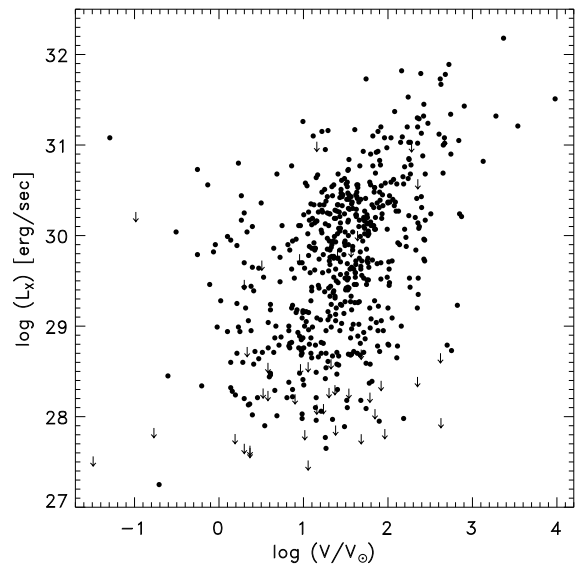, width=8cm} \end{center}
  \caption{X-ray luminosity $\log L_x$ (erg/s) plotted against stellar volume
  ($V_\odot$) for nearly 600 well-characterized Orion stars.  The plusses are
  quiescent levels and dots are time-averaged levels for COUP sources while arrows
  show X-ray nondetections.  Preibisch et al., in preparation.  }
  \end{figure}

On the other hand, one might emphasize the absence of a $L_x-P_{rot}$ relation and
the strength of the $L_x-$Mass$-$ Volume relationships.  The 5 orders of magnitude
range in $L_x$ is largely attributable to the relationships.  Magnetic activity
dependencies on mass and volume in main sequence stars are known but are much
weaker than seen in the PMS population.  The importance of volume and unimportance
of rotation is qualitatively expected from some models of distributed turbulent
dynamos, where the magnetic field is both generated and transported in the
convection zone.  Indeed, it seems difficult to understand how a solar-type dynamo
could possibly operate in a fully convective star where no tachocline or
differential rotational shear layer is expected.

The difficulty in adjudicating this debate is the absence of quantitative
predictions of dynamo models for fully convective stars.  While general
dependencies between surface activity and interior parameters can be established
for solar-type $\alpha-\Omega$ models (Montesinos et al.\ 2001), the models are not
strictly predictive.  The state of distributed dynamo models is even more primitive
despite some recent self-consistent calculations (e.g.  K\"uker \& R\"udiger 1997,
Kitchatinov \& R\"udiger 1999, K\"uker \& Stix 2001).  Efforts must be made to
establish predicted relationships between magnetic field generation and the
principal global parameters -- rotation, mass and volume -- for different classes
of dynamo mechanisms.

\begin{figure}[ht] \begin{center} \epsfig{file=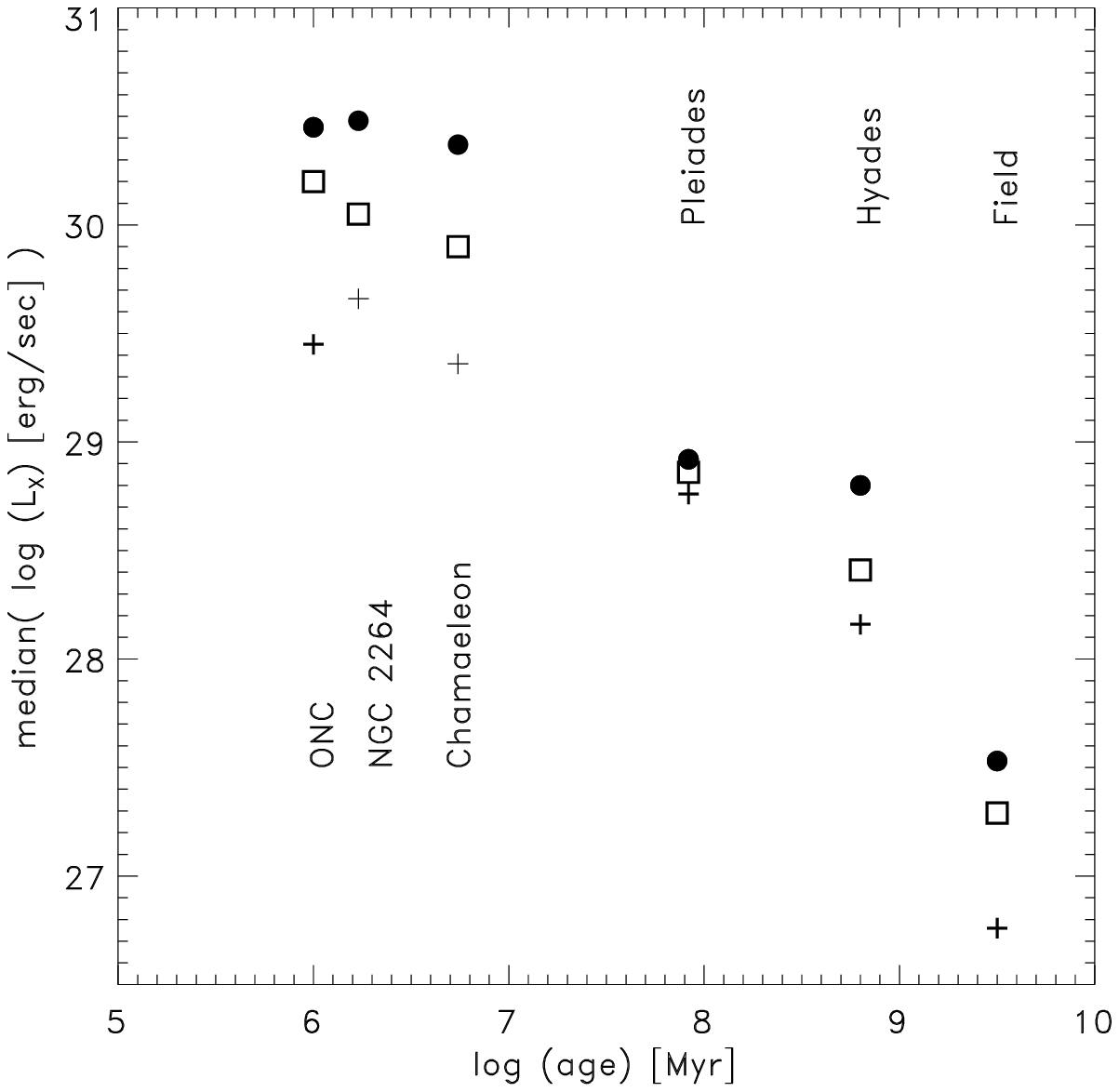, width=7cm}
  \epsfig{file=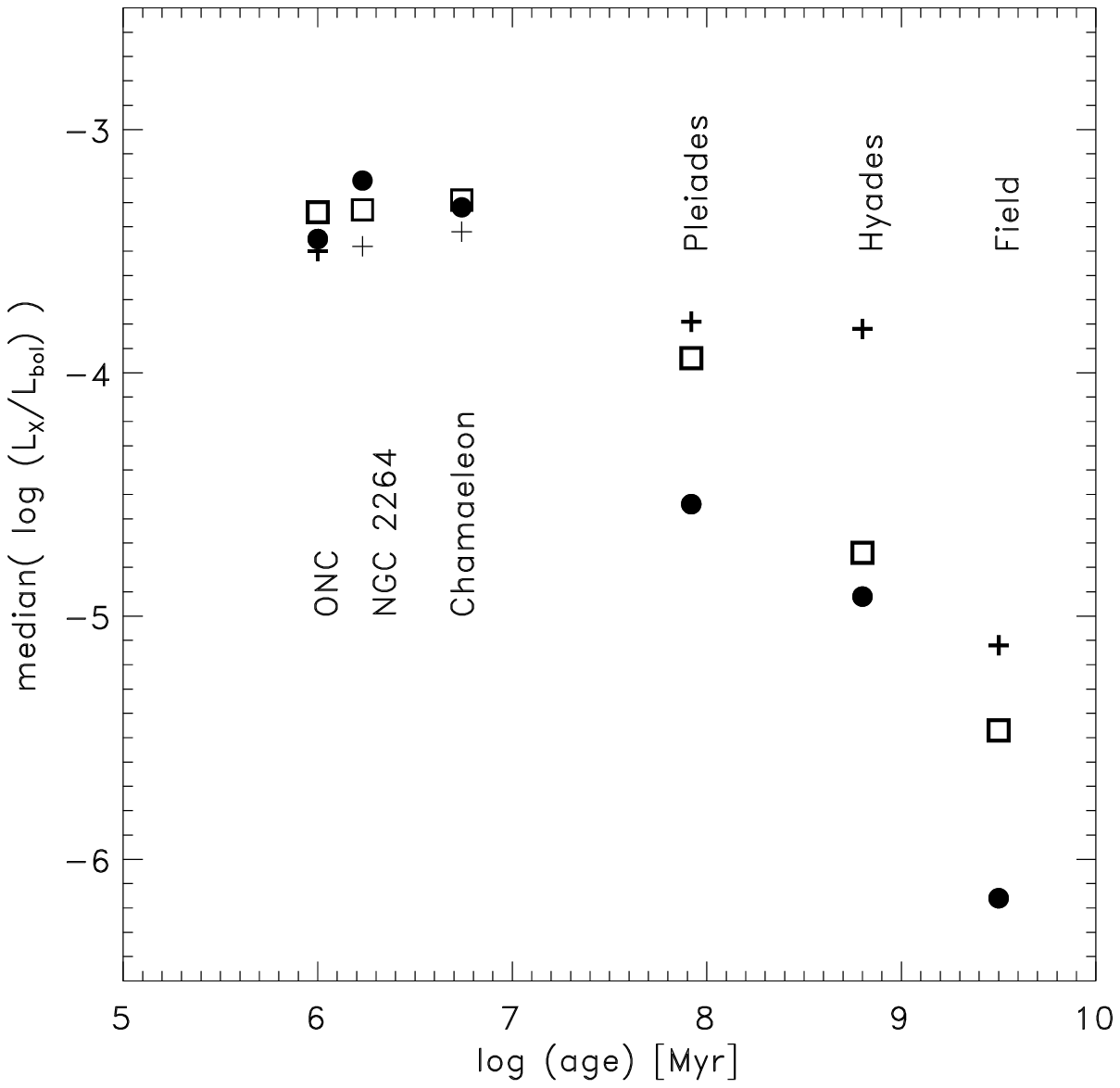, width=8cm} \end{center} \caption{Median X-ray
  luminosity $L_x$ (top) and emissivity $L_x/L_{bol}$ as a function of stellar age
  for lightly absorbed stars in the COUP study of the ONC in comparison with main
  sequence stars of various ages.  The Pleiades, Hyades and solar neighborhood field
  star results are obtained from $ROSAT$ studies.  The solid dots show the
  $0.9\!-\!1.2\,M_\odot$ (G-type on the main sequence) stars, squares show the
  $0.5\!-\!0.9\,M_\odot$ (K-type on the main sequence) stars, and crosses the
  $0.1\!-\!0.5\,M_\odot$ (M-type on the main sequence) stars.  Preibisch et al., in
  preparation.  } \end{figure}

Simon (1992) argued that the activity-age relation was not physically fundamental in
main sequence stars, but rather a derivative effect from activity-rotation and
rotation-age relations.  In contrast, we find in COUP studies an activity-age
relation in PMS stars even in the absence of an activity-rotation relation.  Figure
3 shows the evolution of both the X-ray luminosity $\log L_x$ and emmisivity $\log
L_x/L_{bol}$ for stellar ages ranging from $\log t \simeq 6.0$ to 9.5 yr.  Again we
emphasize that no statistical biases should be present in the COUP study:  the
sample is complete in both in its stellar sample and in the X-ray detections, and
large enough that mass-stratified subsamples can be independently examined.

Taken together, the COUP results in Figures $1-3$ are quite different from those
seen in main sequence stars.  One possible explanation is that the magnetic dynamo
mechanism, whatever it may be, is saturated in the stellar volume or surface.  This
would account for a drop in surface activity as the star descends the convective
Hayashi tracks that is uncorrelated with rotation.  Alternative explanations would
be based on changes in interior structure, such as rotational coupling between
surface and interior layers or the growth of a radiative core in older and more
massive PMS stars.

\section{Magnetic flares and the protoplanetary disk}

The circumstellar disks around PMS stars where planetary systems form are
generally considered to consist of cool, neutral molecular material in
thermodynamic equilibrium.  With characteristic temperatures ranging from 20 to
1500 K, they emit primarily in the infrared and submillimeter bands.  Except for
stars residing close to O stars whose ultraviolet radiation can destroy disks by
photoevaporation (Hollenbach et al.\ 2000), protoplanetary disks have, until the
recent studies outlined here, been treated as closed and isolated structures.

However, there are long-standing characteristics of ancient meteorites which are
exceedingly difficult to explain within the context of a solar nebula in
thermodynamic equilibrium.  These include the enormous quantities of flash melted
chondrules, the calcium-aluminum-rich inclusions (CAIs) with strange combinations
of short-lived radioisotopic anomalies, and the gas-rich grains with spallogenic
$^{21}$Ne excesses correlated with energetic particle track densities (see reviews
by Jones et al.\ 2000, Goswami \& Vanhala 2000, and Woolum \& Hohenberg 1993).
These findings still are deeply disturbing as no single, straightforward
astrophysical explanation can account for all of them.  Some problems seem to be
most readily explained by the injection of recently synthesized radionuclides from
supernovae or other stellar ejecta into the solar nebula, while others seem to
require spallation by MeV particles within the nebula.  Other phenomena are often
attributed to heating by shocks from supersonic bodies in the nebula.

\begin{figure}[ht] \begin{center}{\bf See full version at 
  http://www.astro.psu.edu/users/edf/CS13.pdf} 
  \end{center} \caption{Cartoon illustrating the geometry of magnetic fields and
  high energy photons and particles with respect to the protoplanetary disk.
  (Feigelson 2003) } \end{figure}
  
X-ray astronomical studies of PMS stars have a powerful role in developing models
for these meteoritic mysteries.  They show that magnetic reconnection flares,
orders of magnitude more powerful and frequent than the strongest flares seen in
the contemporary Sun, are ubiquitous in solar mass stars throughout the PMS phases
of evolution.  After Einstein Observatory studies of PMS X-rays, I suggested a link
between X-ray flares and meteoritic issues (Feigelson 1982) and argued that `{\it
invoking high levels of energetic particle or radiation fluxes associated with
magnetic activity in the early Sun is no longer ad hoc}' (Feigelson et al.\ 1991).
After Chandra studies like COUP, this statement can be strengthened:  {\it
Consequences of high energy radiation, both photons and particles, associated with
frequent and powerful magnetic reconnection events within PMS systems cannot be
avoided}.  The main uncertainty in applying this principle is the geometrical
relationship between the flaring field lines and disk region of interest (Feigelson
et al.\ 2002).  Figure 4 provides a visual framework for these issues with
a reasonable guess concerning the geometry.

The question is no longer whether high energy photons and particles, far out of
thermodynamic equilibrium with the thermal material, are generated in PMS systems.
The questions are rather:  What are the geometrical configurations of the
reconnecting fields?  and What are the consequent effects on disk gases and solids?
Feigelson et al.\ (2002) gives a Chandra-based estimate of the high energy particle
irradiation of the disk and discusses the uncertain magnetic geometry.
Implications for the meteoritic anomalies of shortlived radionuclides are discussed
by Leya et al.\ (2003), Gounelle et al.\ (2004), Marhas \& Goswami (2004), Desch et
al.\ (2004) and references therein.

Here, I briefly review recent investigations of the possible effects of X-ray
(rather than MeV particle) irradiation of the disk.

\section{Disk ionization sources}

The first discussion of protoplanetary disk ionization by Gammie (1996) cited
Galactic cosmic rays as the likely penetrating agent.  In standard structure
models, ionization would penetrate all the way to the midplane in the outer regions
of the disk while a midplane neutral `dead zone' would prevail in the inner disk.
One concern is whe\-ther the more numerous low energy cosmic rays can penetrate the
tangled magnetic fields of the surrounding molecular cloud and the flow of
partially ionized outflows from the young stellar system (analogous to the solar
Forbush effect).  Desch et al.\ (2004) considers these problems and find that
cosmic ray penetration is likely quite efficient.

An analogous treatment of disk ionization using PMS X-rays was developed by
Glassgold et al.\ (1997) and Igea \& Glassgold (1999) who showed that Compton
processes increase X-ray penetration through column densities of order
$10^{24}-10^{25}$ cm$^{-3}$ (see review by Glassgold et al.\ 2000).  As cosmic ray
fluxes should be roughly spatially uniform but PMS X-rays fluxes increase as
$r^{-2}$ or faster (due to absorption of the softer photons) as one approaches a
young star, X-ray ionization will always exceed cosmic ray ionization sufficiently
close to a star.  The cross-over distance for these two ionization sources is
model-dependent but, for PMS solar levels around $\log L_x \simeq 30$ erg/s and
characteric plasma temperatures of $kT \simeq 1-5$ keV, likely occurs around a few
AU (Matsumura \& Pudritz 2003).

Collisional ionization of atoms with low first ionization potentials, such as
potassium and sodium, may also be important when disk temperatures exceed $\simeq
800$ K.  Differing treatments of these metal ions leads to considerable differences
in the inferred steady-state ionization level of the disk, even for constant
external ionization rates (Fromang et al.\ 2002).  The theory of the X-ray
ionization {\it rate} appears satisfactory, but calculations of the X-ray
ionization {\it fraction} depend on poorly established recombination rates.

X-ray ionization effects are only one of several contributors to the very
complex and nonlinear interplay between the thermodynamics, dynamics (including
magnetic field effects), gas-phase chemistry and gas-grain interactions in
protoplanetary disks.  Astronomical observations of molecular species are
rapidly emerging, but suffer from insufficient spatial resolution and
sensitivity and often give confusing results.  Given the complexities of disk
thermodynamics and chemical evolution, confident understanding of X-ray chemical
effects may await the ALMA interferometer which will image many molecules across
the disks of nearby PMS stars.

\section{Disk turbulence and its implications}

Provided a weak magnetic field was entrained in the disk during collapse, the
Keplerian shear in the ionized zone interacts with the field to induce the
magnetorotational instability, also known as the Balbus-Hawley instability, which
quickly develops into a full spectrum of MHD turbulence including both vertical and
radial mixing.  Ionization fractions as low as $10^{-11}$ may give adequate
coupling which is achieved when the collision frequency of neutrals with ions
exceeds the local epicyclic frequency (Blaes \& Balbus 1994).  Recent studies have
investigated the growth of the instability including effects of vertical
stratification, finite conductivity, ohmic dissipation, ambipolar diffusion and
Hall effects (Salmeron \& Wardle 2003; Desch 2004).

This ionization-induced turbulence may be of considerable importance for planet
formation and dynamical evolution in several ways:  \begin{enumerate}

\item The radial viscosity associated with the active turbulent zone may cause the
flow of material from the outer disk into the inner disk, and thereby into the
bipolar outflows and onto the protostar.  This may solve a long-standing problem in
young stellar studies:  a completely neutral disk should have negligible viscosity
and thus cannot efficiently be an accretion disk.  Armitage et al.\ (2001) suggest
that a long-term disk instability may be also occur:  when the dead zone
temperature exceeds $\simeq 800$ K so that collisional ionization rises, the entire
inner disk may suddenly become viscous and spiral towards the star.  Such bursts of
accretion may give rise to the FU Orionis phenomena and episodic accretion seen as
symmetric shocks in large-scale Herbig-Haro outflows.

\item Ionization-induced turbulence will produce density inhomogeneities which can
overwhelm other extensively studied dynamical structures -- such as spiral
instabilities, gaps and streams -- that emerge from dynamical interactions between
protoplanets and gaseous disks.  A possible crucial consequence may be that
turbulent density perturbations will affect gap formation and produce stochastic
radial torques on the protoplanet.  The protoplanet's radial motion may then
resembles a random walk rather than the rapid inspiralling called Type I migration.
Several groups have made MHD studies of this process (Nelson \& Papaloizou 2004;
Papaloizou \& Nelson 2003; Winters et al.  2003; Menou \& Goodman 2004; Laughlin
et al.\ 2004; Hersant et al.\ 2004).  Results from one of these calculations is
illustrated in Figure 5.

\end{enumerate}

\begin{figure}[ht] \begin{center} 
  {\bf See full version at http://www.astro.psu.edu/users/edf/CS13.pdf}
  \end{center} \caption{Three-dimensional
  magnetohydrodynamical simulation of a disk with a protoplanet.  The top panel
  shows a neutral disk with spiral instabilities and streams which cause Type I
  migration.  The bottom panel shows the same model with turbulent inhomogeneities
  caused by instabilities arising from ionization. (Laughlin et al.\ 2004) } \end{figure}

It is thus possible that X-ray emission plays an important role in regulating the
structure and dynamics of planetary systems.  Weaker PMS X-ray sources would
produce disks with weaker turbulence, stronger Type I migration with rapid
inspiralling, resulting in planetary systems with `hot Jupiters' like 51 Peg.
Stronger PMS X-ray sources might produce disks with stronger turbulence, random
walk radial motion so that Jovian protoplanets remain in the mid-disk regions; the
resulting planetary systems may resemble those around $\upsilon$ And and our Sun.
Random walk migration weakens disk gaps and speeds up protoplanetary accretion.
This may account for the much higher masses of active zone Jovian planets compared
to dead zone terrestrial planets (Matsumura \& Pudritz 2003).  Jovian planets
undergoing turbulent random walks also accrete their full masses more rapidly than
those in non-turbulent disks, which can alleviate planet formation timescale
problems in some earlier models (Rice \& Armitage 2003).

Finally, we note that PMS X-rays are a major sources of ionization at the base on
outflows from protostellar disks which produce the emission line Herbig-Haro
objects and molecular bipolar outflows (Shang et al.  2002, Ferro-Font\'an et al.
2003).  This is a profound result:  if low-mass PMS stars were not magnetically
active and profusely emitting penetrating photoionizing radiation, then the
coupling between the Keplerian orbits in the disk and the magnetocentrifugally
orbits spiralling outward perpendicular to the disks might be much less efficient
than we see.  Without powerful disk outflows, molecular cloud energetics, disk
evolution and even planet formation would be considerably altered.

\section{Other X-ray effects on disks}

It has been known for 20 years that X-ray irradiation of a molecular environment
will induce a complex series of {\bf molecular-ion and radical chemical
reactions}.  This was first applied to protoplanetary disk chemistry by Ai\-kawa
\& Herbst (1999) and further studied by Aikawa \& Herbst (2001) Markwick et al.\
(2002) and Semenov et al.\ (2004).  The latter authors find that, from the
chemical perspective, X-ray ionization is a critical ionizing source in the
surface disk layers close to the star ($r<10$ AU) and also at deeper
intermediate layers throughout the disk where it drives a rich ion-molecular
chemistry (Figure 6).  CN, HCO+ and C$_2$H abundances may be good tracers of
photoionization effects, though it is often difficult to distinguish X-ray and
ultraviolet irradiation from global disk observations.  The CN/HCN ratio is
elevated in the disks around X-ray luminous T Tauri and Herbig Ae/Be stars
pointing to photochemistry, but again the roles of UV and X-ray effects are
again intertwined (Thi et al.\ 2004).  X-ray heating should also lead to ice
evaporation and enhanced gaseous abundances of molec ules such as methanol,
though this has not yet been seen.  X-ray ionization may also be manifested in
the chemistry of infalling envelopes around Class 0 protostars (Doty et al.\
2004).  However, evidence for X-ray ionization-induced chemical species is often
mixed; for instance, Ceccarelli et al.\ (2004) find H$_2$D$^+$ abundance in the
TW Hya disk is consistent with cosmic ray ionization alone.

X-ray absorption contribute to the {\bf warming of outer molecular layers of the
disk}.  Glassgold et al.\ (2004) show that, in the outermost layer, the gas is
heated to 5000 K, far above the equilibrium dust temperature.  One consequence is
a thick layer of warm CO responsible for the strong CO infrared bands seen from
several young disks.  However, the relative importance of heating from X-ray
irradiation and mechanical heating from external wind-disk interaction or shock
dissipation of internal turbulence is unclear.  Gorti \& Hollenbach (2004) model
the heating to partially dissipated older T Tauri disks.  X-ray heating is
important in the outer disk layers and the disk edge assumed to lie 0.1 AU from
the star.  They also predict a correlation between the neutral oxygen 63$\mu$m
line strength and X-ray irradiation.  X-ray excitation of disk molecules is a
plausible origin for the ro-vibrational lines of H$_2$ seen in several T Tauri
stars including, surprisingly, one classified as a disk-free weak-lined T Tauri
star (Weintraub et al.\ 2000, Bary et al.\ 2003).

\begin{figure}[ht] \begin{center} \epsfig{file=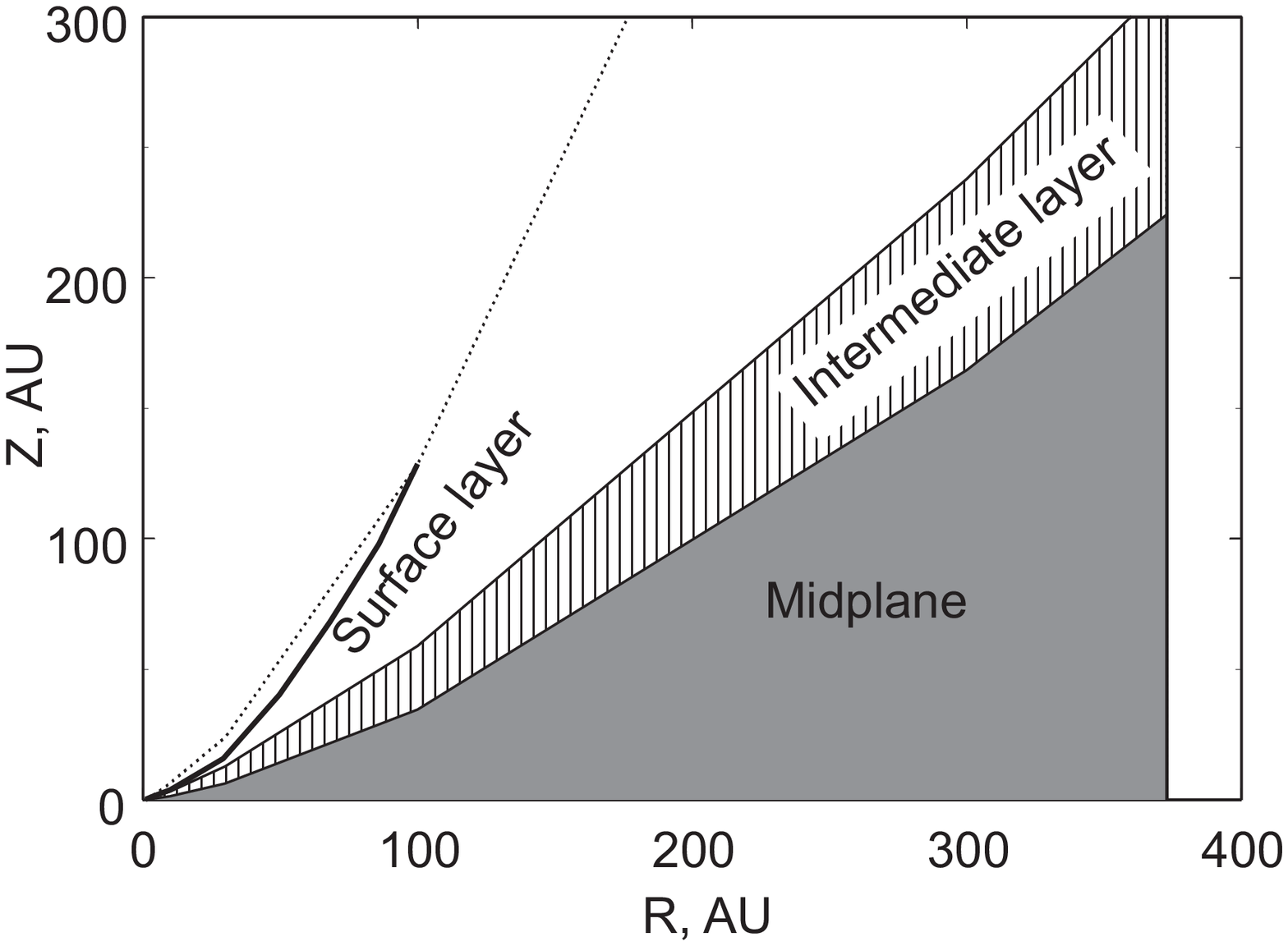, width=7cm}
  \epsfig{file=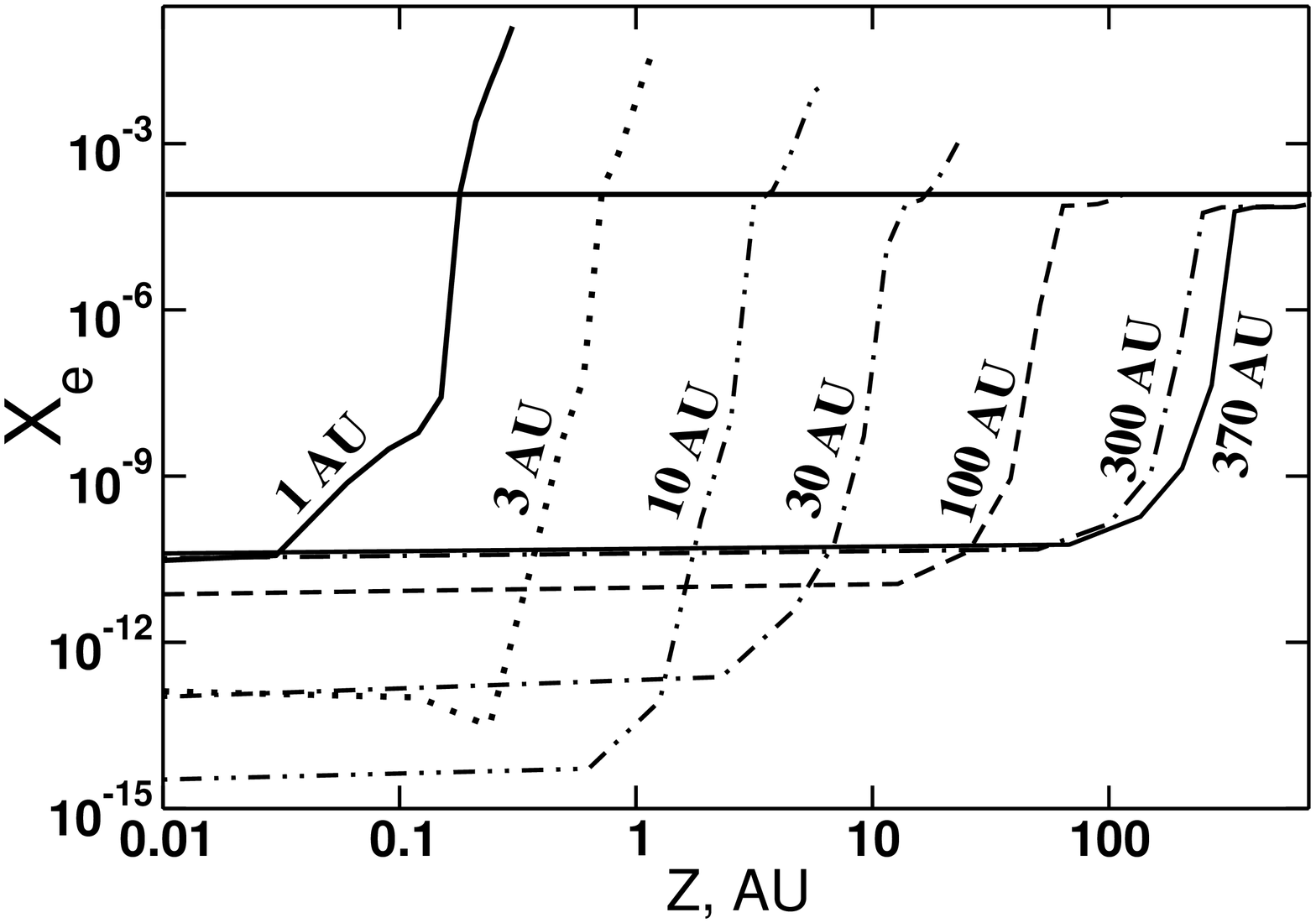, width=7cm} \end{center} \caption{Disk structure (top)
  and ionization fraction (bottom) resulting from a protoplanetary disk model with
  a complex ion-molecular chemical reaction network.  The solid line defining the
  top of the outer layer denotes the $10^{-4}$ ionization level.  (Semenov et al.\
  2004) } \end{figure}
  
When PMS {\bf X-rays are absorbed by dust grains}, the localized heating may desorb
molecules in the grain mantle and increasing some molecular abundances in the gas
phase (Najita et al.  2001).  Conceivably, such spot heating may produce liquid
water which would change mineral chemistry; this is a suggested explanation for an
unexpected infrared emission band attributed to calcite (Ceccarelli et al.  2002).

Another X-ray heating effect in the disk may be the {\bf flash melting of dustballs
producing meteoritic} {\bf chondrules and/or CAIs} (Shu et al.\ 2001).  Such models are
attractive in their integrated treatment of spallogenic isotopic anomalies and
flash melting.  But they are challenged by the need to transport the mm-cm-sized
melted rocks from the vicinity of the flares near the star to the Asteroid Belt
where they are now prevalent.

\section{Final comments}

The COUP study is solidifying our knowledge of the {\it phenomenology} of X-ray
emission and magnetic activity in PMS stars.  It confirms earlier Chandra findings
that the age-activity-rotation relationships seen in main sequence late-type stars
do not apply in the earlier phases of evolution.  The main predictors of stellar
activity is a weave of stellar mass-luminosity-volume variables, not rotation.
However, our {\it understanding} of this finding is not reaching clarity.  We do
not know whether to treat the effect as evidence for a variant of standard
$\alpha-\Omega$ dynamos (i.e., the supersaturated regime) or for a different dynamo
mechanism (i.e., a distributed turbulent convective dynamo).

This is only one of many avenues of research being conducted with COUP, and COUP is
only one of many efforts in studying PMS stars using Chandra.  From a purely
astronomical viewpoint, Chandra will emerge with catalogs of tens of thousands of
PMS stars, most previously unstudied, in many star formation regions at distances
of $0.5-10$ kpc in the Galactic disk.  This alone can give tremendous impetus to
the study of PMS stars in the coming years.

Observational constraints on the various aspects of X-ray irradiation of
protoplanetary disks are rapidly emerging.  Impressive advances in infrared
instrumentation have been achieved with the Spitzer Space Telescope and
ground-based detectors such as Visir on the ESO Very Large Telescopes and T-ReCS on
the Gemini Telescope.  Some insights will await the high resolution and sensitivity
of ALMA in the submillimeter band.

On the X-ray front, COUP is already providing two lines of evidence that disk
irradiation by X-rays is common.  First, COUP detects 2/3 of Orion stars with disks
imaged with the Hubble Space Telescope (J.\ Kastner et al., in
preparation).  In a few cases, X-ray absorption will measure the line-of-sight
column density through the disk.  Second, in a few of the brightest X-ray sources
associated with very young PMS stars, we detect the 6.4 keV fluorescent emission
line (M.\ Tsujimoto et al., in preparation).  We argue this arises from reflection
off of the cold disk.  The next generation of X-ray telescopes, unfortunately over
a decade in the future, conceivably could infer the structure of the disks through
reverberation mapping of X-ray flares.

\begin{acknowledgements} This work was supported by Chandra Guest Observer grants
GO3-4009A supporting COUP and ~ GO4-5505X supporting ~ study of older disks.  Thomas
Preibisch (MPIfR) kindly shared COUP results prior to publication. Dmitry Semenov 
(MPIA), Gregory Laughlin (UC Santa Cruz) and Al Glassgold (UC Berkeley) 
provided valuable assistance on the manuscript.
\end{acknowledgements}


\begin{thebibliography}{}

\bibitem[]{} Aikawa, Y.~\& Herbst, E.\ 1999, AA, 351, 233

\bibitem[]{} Aikawa, Y.~\& Herbst, E.\ 2001, AA, 371, 1107

\bibitem[]{} Armitage, P.~J., Livio, M., \& Pringle, J.~E.\ 2001, 
MNRAS, 324, 705 

\bibitem[]{} Barnes, S.~A.\ 2003, ApJ, 586, 464

\bibitem[]{} Bary, J.~S., Weintraub, D.~A., \& Kastner, J.~H.\ 2003, ApJ, 586, 1136

\bibitem[]{} Blaes, O.~M.~\& Balbus, S.~A.\ 1994, ApJ, 421, 163

\bibitem[]{} Ceccarelli, C., Caux, E., Tielens, A.~G.~G.~M., Kemper, F., Waters,
L.~B.~F.~M., \& Phillips, T.\ 2002, AA, 395, L29

\bibitem[]{} Ceccarelli, C., Dominik, C., Lefloch, B., Caselli, P., \& Caux, E.\
2004, ApJL, 607, L51

\bibitem[]{} Desch, S.~J., Connolly, H.~C., \& Srinivasan, G.\ 2004, ApJ, 602, 528

\bibitem[]{} Desch, S.~J.\ 2004, ApJ, 608, 509

\bibitem[]{} Doty, S.~D., Sch{\" o}ier, F.~L., \& van Dishoeck, E.~F.\ 2004, AA,
418, 1021

\bibitem[]{} Feigelson, E.~D.\ 1982, Icarus, 51, 155

\bibitem[]{} Feigelson, E.~D., Giampapa, M.~S., \& Vrba, F.~J.\ 1991, in {\it The
Sun in Time}, 658

\bibitem[]{} Feigelson, E.~D.~\& Montmerle, T.\ 1999, ARAA, 37, 363

\bibitem[]{} Feigelson, E.~D., Broos, P., Gaffney, J.~A., et al.\ 2002, ApJ, 574,
258

\bibitem[]{} Feigelson, E.\ 2003, in {\it Stars as Suns:  Activity, Evolution and
Planets}, IAU Symp 219

\bibitem[]{} Feigelson, E.~D., Gaffney, J.~A., Garmire, G., Hillenbrand, L.~A., \&
Townsley, L.\ 2003, ApJ, 584, 911

\bibitem[]{} Feigelson, E.~D., Hornschemeier, A.~E., Micela, G.  et al.\, 2004,
ApJ, in press

\bibitem[]{} Ferro-Font{\' a}n, C.~\& de Castro, A.~I.~G.\ 2003, MNRAS, 342, 427

\bibitem[]{} Flaccomio, E., Damiani, F., Micela, G., et al.\ 2003a, ApJ, 582, 398

\bibitem[]{} Flaccomio, E., Micela, G., \& Sciortino, S.\ 2003b, As\&Ap, 402, 277

\bibitem[]{} Gammie, C.~F.\ 1996, ApJ, 457, 355

\bibitem[]{} Getman, K.~V., Flaccomio, E., Broos, P.~S., et al.\ 2005, ApJS, in
press

\bibitem[]{} Gilman, P.~A.\ 1981, in {\it Cool Stars, Stellar Systems, and the
Sun}, 2, 165

\bibitem[]{} Glassgold, A.~E., Feigelson, E.~D., \& Montmerle, T.\ 2000, in {\it
Protostars and Planets IV}, 42

\bibitem[]{} Glassgold, A.~E., Najita, J.\ \& Igea, J.\ 2004, ApJ, in press

\bibitem[]{} Gorti, U.\ \& Hollenbach, D.~J.\ 2004, ApJ, in press

\bibitem[]{} Goswami, J.~N.~\& Vanhala, H.~A.~T.\ 2000, in {\it Protostars and
Planets IV}, 963

\bibitem[]{} Gounelle, M., Shu, F.~H., Shang, H., et al.\ 2004, Lunar Plan Inst
Conf, 35, 1829

\bibitem[]{} G\"udel, M., Guinan, E.~F., \& Skinner, S.~L.\ 1997, ApJ, 483, 947

\bibitem[]{} Gunn, A.~G., Mitrou, C.~K., \& Doyle, J.~G.\ 1998, in {\it Cool Stars,
Stellar Systems, and the Sun}, 10, 1257

\bibitem[]{} Herbst W., Bailer-Jones, C.~A., Mundt, R., Meisenheimer, K.\ \&
Wackermann, R.\ 2002, AA, 396, 513

\bibitem[]{} Hersant, F., Dubrulle, B.  \& Hure, J.~M.\ 2004, AA, in press

\bibitem[]{} Hollenbach, D.~J., Yorke, H.~W., \& Johnstone, D.\ 2000, in {\it
Protostars and Planets IV}, 401

\bibitem[]{} Jones, R.~H., Lee, T., Connolly, H.~C., Love, S.~G., \& Shang, H.\
2000, in {\it Protostars and Planets IV}, 927

\bibitem[]{} Kitchatinov, L.~L.~\& R{\" u}diger, G.\ 1999, AA, 344, 911

\bibitem[]{} K\"uker, M.~\& R\"udiger, G.\ 1997, AA, 328, 253

\bibitem[]{} K{\" u}ker, M.~\& Stix, M.\ 2001, AA, 366, 668

\bibitem[]{} Leya, I., Halliday, A.~N., \& Wieler, R.\ 2003, ApJ, 594, 605

\bibitem[]{} Laughlin, G., Steinacker, A., \& Adams, F.~C.\ 2004, ApJ, 608, 489

\bibitem[]{} Marhas, K.~K.~\& Goswami, J.~N.\ 2004, New Astro Rev, 48, 139

\bibitem[]{} Markwick, A.~J., Ilgner, M., Millar, T.~J., \& Henning, T.\ 2002, AA,
385, 632

\bibitem[]{} Matsumura, S.~\& Pudritz, R.~E.\ 2003, ApJ, 598, 645

\bibitem[]{} Menou, K.~\& Goodman, J.\ 2004, ApJ, 606, 520

\bibitem[]{} Mohanty, S.~\& Basri, G.\ 2003, in {\it Cool Stars, Stellar Systems,
and the Sun}, 12, 683

\bibitem[]{} Montesinos, B., Thomas, J.~H., Ventura, P., \& Mazzitelli, I.\ 2001,
MNRAS, 326, 877

\bibitem[]{} Najita, J., Bergin, E.~A., \& Ullom, J.~N.\ 2001, ApJ, 561, 880

\bibitem[]{} Nelson, R.~P.~\& Papaloizou, J.~C.~B.\ 2004, MNRAS, 350, 849

\bibitem[]{} Papaloizou, J.~C.~B.~\& Nelson, R.~P.\ 2003, MNRAS, 339, 983

\bibitem[]{} Pizzolato, N., Maggio, A., Micela, G., Sciortino, S., \& Ventura, P.\
2003, AA, 397, 147

\bibitem[]{} Rice, W.~K.~M.~\& Armitage, P.~J.\ 2003, ApJL, 598, L55

\bibitem[]{} Salmeron, R.~\& Wardle, M.\ 2003, MNRAS, 345, 992

\bibitem[]{} Schatzman, E.\ 1962, Ann d'Astrophys, 25, 18

\bibitem[]{} Schrijver, C.~J.~\& Zwaan, C.\ 2000, {\it Solar and Stellar Magnetic
Activity}, NY:Cambridge University Press

\bibitem[]{} Schulz, N.~S., Canizares, C., Huenemoerder, D., et al.\ 2001, ApJ,
549, 441

\bibitem[]{} Semenov, D., Wiebe, D., \& Henning, T.\ 2004, AA, 417, 93

\bibitem[]{} Shang, H., Glassgold, A.~E., Shu, F.~H., \& Lizano, S.\ 2002, ApJ,
564, 853

\bibitem[]{} Shu, F.~H., Shang, H., Gounelle, M., Glassgold, A.~E., \& Lee, T.\
2001, ApJ, 548, 1029

\bibitem[]{} Simon, T.\ 1992, in {\it Cool Stars, Stellar Systems, and the Sun}, 7,
3

\bibitem[]{} Skumanich, A.\ 1972, ApJ, 171, 565

\bibitem[]{} Stassun, K.~G., Ardila, D.~R., Barsony, M., Basri, G., \& Mathieu,
R.~D.\ 2004, AJ, 127, 3537

\bibitem[]{} Thi, W.-F., van Zadelhoff, G.-J.\ \& van Dishoeck, E.\~F.  2004, AA,
in press

\bibitem[]{} Valenti, J.~A., Johns-Krull, C.~M., \& Piskunov, N.\ 1998, in {\it
Cool Stars, Stellar Systems, and the Sun}, 10, 1357

\bibitem[]{} Weintraub, D.~A., Kastner, J.~H., \& Bary, J.~S.\ 2000, ApJ, 541, 767

\bibitem[]{} Wilson, O.~C.\ 1963, ApJ, 138, 832

\bibitem[]{} Winters, W.~F., Balbus, S.~A., \& Hawley, J.~F.\ 2003, ApJ, 589, 543

\bibitem[]{} Woolum, D.~S.~\& Hohenberg, C.\ 1993, in {\it Protostars and Planets
III}, 903

\end{thebibliography}
\end{document}